\documentclass[final,3p,times,fleqn]{elsarticle}
\usepackage{mathtools,booktabs}
\usepackage{bm}
\usepackage{autobreak}
\usepackage{amsmath}
\usepackage{amsthm}
\usepackage{amssymb}
\usepackage{mathrsfs}
\usepackage{siunitx}
\usepackage{gensymb}
\usepackage{subfigure}
\usepackage{epstopdf}
\usepackage{color}
\usepackage[svgnames]{xcolor}
\usepackage{hyperref}
\usepackage{numprint}

\biboptions{sort&compress}
\journal{}
\begin{document}
	\begin{frontmatter}
		\title{Phase-field-based lattice Boltzmann method for the transport of insoluble surfactant in two-phase flows}
		\author[a,b,c]{Chengjie Zhan}
		%\ead{zhancj@hust.edu.cn}
		\author[d]{Hong Liang}
		%\ead{lianghongstefanie@163.com}
		\author[a,b,c]{Zhenhua Chai \corref{cor1}}
		\ead{hustczh@hust.edu.cn}
		\author[a,b,c]{Baochang Shi}
		%\ead{shibc@hust.edu.cn}
		\address[a]{School of Mathematics and Statistics, Huazhong University of Science and Technology, Wuhan 430074, China}
		\address[b]{Institute of Interdisciplinary Research for Mathematics and Applied Science, Huazhong University of Science and Technology, Wuhan 430074, China}
		\address[c]{Hubei Key Laboratory of Engineering Modeling and Scientific Computing, Huazhong University of Science and Technology, Wuhan 430074, China}
		\address[d]{Department of Physics, Hangzhou Dianzi University, Hangzhou 310018, China}
		\cortext[cor1]{Corresponding author.}
		\begin{abstract} 
			In this work, we present a general second-order phase-field model for the transport of insoluble surfactant in incompressible two-phase flows. In this model, the second-order local Allen-Cahn equation is applied for interface capturing, a general form of the simple scalar transport equation [S. S. Jain, J. Comput. Phys. 515, 113277 (2024)] is adopted for interface-confined surfactant, and the consistent and conservative Navier-Stokes equations with the Marangoni force is used for fluid flows. To solve this model, we further developed a mesoscopic lattice Boltzmann (LB) method, in which the LB model for surfactant transport equation is proposed under the general LB framework for the convection-diffusion type equation, and it can correctly recover the governing equation for surfactant transport. The accuracy of the present LB method is tested by several benchmark problems, and the numerical results show it has a good performance for the transport of the insoluble surfactant in two-phase flows. 
		\end{abstract}
	\end{frontmatter}		
\section{Introduction}
The surfactants can be used to reduce the surface tension and generate the Marangoni force of the multiphase system, and has been widely used to control the dynamics of multiphase flows in many different fields, for instances, enhanced oil recovery \cite{Sun2014EF}, droplet manipulation in microfluidics \cite{Eggleton2001PRL,Booty2005JFM,Baret2012LC}, and froth flotation \cite{Li2012WST}, to name but a few. 
Generally, the surfactants can be divided into two categories, the soluble form and the insoluble form, according to whether they are only adsorbed at the fluid interface. 
On the one hand, for the soluble surfactants in two-phase flows, a total free energy functional has been established and the Cahn-Hilliard-based diffuse-interface framework is used to model the fluid interface and the surfactant \cite{Teigen2009CMS,Teigen2011JCP}. 
On the other hand, for insoluble surfactants, Stone \cite{Stone1990PF} proposed a simple derivation of the governing equation for the surfactant transport along a deforming interface,
\begin{equation}
	\frac{\partial\hat{\psi}}{\partial t}+\nabla_s\cdot\left(\hat{\psi}\mathbf{u}\right)=\nabla_s\cdot D\nabla_s\hat{\psi},
\end{equation}
where the surfactant concentration $\hat{\psi}$ (amount of scalar per unit area of the interface) is only defined on the interface. $\mathbf{u}$ is the fluid velocity, $D$ is the diffusivity, and $\nabla_s$ represents the surface gradient. This sharp-interface model was extended by adopting the definition of surfactant into the neighborhood of the interface in Ref. \cite{Xu2003JSC}. Then it was further reformulated in the entire fluid domain by introducing a delta function \cite{Teigen2011JCP},
\begin{equation}\label{eq-Teigen}
	\frac{\partial\left(\delta\hat{\psi}\right)}{\partial t}+\nabla\cdot\left(\delta\hat{\psi}\mathbf{u}\right)=\nabla\cdot\delta D\nabla\hat{\psi},
\end{equation}
where $\delta$ is the delta function of the interface $\Gamma$ in fluid domain $\Omega$, and satisfies $\int_{\Gamma}\hat{\psi}\mathrm{d}\Gamma=\int_{\Omega}\delta\hat{\psi}\mathrm{d}\Omega$.
In this model, the governing equation is defined in the entire fluid domain by the Eulerian representation, which is easy to be solved by a simple numerical method, such as the finite-difference methods \cite{Teigen2011JCP,Liu2018JFM}. 
Recently, Hu \cite{Hu2021AML} proposed an improved form of the above model,
\begin{equation}\label{eq-Hu}
	\frac{\partial\left(\delta\hat{\psi}\right)}{\partial t}+\nabla\cdot\left(\delta\hat{\psi}\mathbf{u}\right)=\nabla\cdot\delta D\left[\mathbf{I}+\left(r-1\right)\mathbf{n}\otimes\mathbf{n}\right]\nabla\hat{\psi},
\end{equation}
which can also be shown in an equivalent form,
\begin{equation}\label{eq-Hu1}
	\frac{\partial\left(\delta\hat{\psi}\right)}{\partial t}+\nabla\cdot\left(\delta\hat{\psi}\mathbf{u}\right)=\nabla\cdot D\left[\nabla\left(\delta\hat{\psi}\right)+\mathbf{J}_s\right],
\end{equation}
where $r\ge0$ is a parameter used to control the diffusion rate in the normal direction, $\mathbf{I}$ and $\mathbf{n}$ are the identity matrix and the interfacial unit normal vector, $\mathbf{J}_s$ is a flux term \cite{Hu2021AML}. 
Moreover, when the fluid interface is captured by the phase-field method with order parameter $\phi$, the fourth-order polynomial of $\phi$ is used to approximate the delta function \cite{Teigen2011JCP,Hu2021AML}. 
However, to compute the surfactant concentration $\hat{\psi}$ in the above models with delta function, the division of $\delta\hat{\psi}$ by $\delta$ is needed, which could cause the numerical stability and/or accuracy issues when $\delta$ goes to zero. 
Additionally, the parameter $r$ has an influence on the numerical results \cite{Hu2021AML}, and how to properly determine its value is also a critical problem.  

Different from above definition of the surfactant concentration in the sharp representation, Jain \cite{Jain2024JCP} constructed a diffuse quantity $\psi$ (concentration per unit volume) in the entire fluid domain, and the integral of $\psi$ in the interface normal coordinates can result in $\hat{\psi}$. Based on this concept, the corresponding transport equation of the insoluble surfactant can be given by  
\begin{equation}\label{eq-Jain0}
	\frac{\partial\psi}{\partial t}+\nabla\cdot\left(\psi\mathbf{u}\right)=\nabla\cdot D\left(\nabla\psi-\psi\frac{\nabla\delta}{\delta}\right),
\end{equation}
which can be derived from Eq. (\ref{eq-Teigen}) or Eq. (\ref{eq-Hu}) ($r=1$) by replacing $\hat{\psi}$ with $\psi$ through the relation $\psi=\delta\hat{\psi}$ \cite{Jain2024JCP}. In addition, the delta function is approximated by $\left|\nabla\phi\right|$, which can be further assumed to be the gradient norm of the equilibrium profile of the phase-field variable, and can be expressed as a second-order polynomial of $\phi$. 
Almost at the same time, Yamashita et al. \cite{Yamashita2024JCP} proposed another similar surfactant transport model based on the phase-field method, in which two extra terms are added in Eq. (\ref{eq-Jain0}) to prevent the numerical diffusion in the normal direction. 
We also note that this model \cite{Yamashita2024JCP} can be derived from Eq. (\ref{eq-Hu}) or (\ref{eq-Hu1}) with $r=2$ and $\psi=\delta\hat{\psi}$.
The distinct advantage of these two diffuse-interface models is that they can overcome the numerical modeling challenge of the surfactant concentration in sharp representation on the Eulerian grid. 

As a diffuse-interface approach, the lattice Boltzmann (LB) method can describe the coupling interactions among different physical fields, and has been widely adopted to study complex flow problems, especially the multiphase flows, due to the features of easy implementation of boundary conditions and fully parallel algorithm \cite{Chen1998ARFM,Aidun2010ARFM}. For the transport of insoluble surfactant, Hu \cite{Hu2021AML} first developed an LB method to solve the transport equation (\ref{eq-Hu1}). However, with the Chapman-Enskog expansion, the governing equation (\ref{eq-Hu1}) cannot be correctly recovered by the LB method since there is an additional term related to the convection term in the recovered macroscopic equation. In addition, a small parameter $\epsilon$ is also introduced in this work to avoid the denominator to be zero when computing the surfactant concentration $\hat{\psi}$, but the value of this small parameter also has an apparent influence on the numerical results. 

In this work, we adopt the second-order diffuse-interface model proposed by Jain \cite{Jain2024JCP} to describe the transport of insoluble surfactant, and extend the model to study two-phase flows with the Marangoni effect. To solve the extended model, a triple-distribution-function mesoscopic LB method is proposed, in which a novel LB model is developed for the transport equation of insoluble surfactant. 
The rest of this work is organized as follows. In Sec. \ref{sec-model}, the governing equations for phase field, surfactant concentration field and flow field are introduced. Then in Sec. \ref{sec-LBM}, the LB method is proposed for different physical fields, followed by numerical validations and discussion in Sec. \ref{sec-simulations}. Finally, some conclusions are summarized in Sec. \ref{sec-conclusion}. 

\section{Mathematical model}\label{sec-model}
In this section, the mathematical model for the transport of insoluble surfactant in two-phase flows will be presented, which is composed of the phase-field model for interface capturing, the diffuse-interface model for the transport of insoluble surfactant, and the incompressible Navier-Stokes equations for fluid flows.
\subsection{Phase-field model for interface capturing}
To capture the dynamics of the fluid interface, we apply the following popular second-order conservative Allen-Cahn equation \cite{Chiu2011JCP},
\begin{equation}\label{eq-ACE}
    \frac{\partial\phi}{\partial t}+\nabla\cdot\left(\phi\mathbf{u}\right)=\nabla\cdot M_{\phi}\left[\nabla\phi-\frac{4\left(\phi_A-\phi\right)\left(\phi-\phi_B\right)}{W\left(\phi_A-\phi_B\right)}\frac{\nabla\phi}{\left|\nabla\phi\right|}\right],
\end{equation}
where $\mathbf{u}$ is the velocity, $M_{\phi}$ is the mobility, and $W$ represents the width of the diffuse interface.
$\phi$ is the order parameter, which is changed smoothly from a constant $\phi_B$ in fluid B to another constant $\phi_A$ ($\phi_A>\phi_B$) in fluid A, and can be used to label the fluid interface $\left(\phi_A+\phi_B\right)/2$. Usually, for one-dimensional problem, the equilibrium profile of the order parameter can be given by \cite{Wang2019Capillarity}
\begin{equation}\label{eq-phieq}
	\phi_{eq}=\frac{\phi_A+\phi_B}{2}+\frac{\phi_A-\phi_B}{2}\tanh\frac{2x}{W},
\end{equation}
where the interface of two phases is assumed at $x=0$.
\subsection{Diffuse-interface model for surfactant transport}
In phase-field context, several definitions of the delta function are available from the literature. Here, based on the previous work \cite{Jain2024JCP}, the delta function $\delta$ and its gradient can be approximated by 
\begin{equation}
    \delta=\left|\nabla\phi\right|=\frac{4\left(\phi_A-\phi\right)\left(\phi-\phi_B\right)}{W\left(\phi_A-\phi_B\right)},\quad\nabla\delta=\frac{4\left(\phi_A+\phi_B-2\phi\right)}{W\left(\phi_A-\phi_B\right)}.
\end{equation}
Then substituting above equation into Eq. (\ref{eq-Jain0}) yields
\begin{equation}\label{eq-Jain}
    \frac{\partial\psi}{\partial t}+\nabla\cdot\left(\psi\mathbf{u}\right)=\nabla\cdot D\left[\nabla \psi-\frac{4\psi\left(\phi_A+\phi_B-2\phi\right)}{W\left(\phi_A-\phi_B\right)}\frac{\nabla\phi}{\left|\nabla\phi\right|}\right].
\end{equation}
Compared to the classic convection-diffusion equation, the second term in the square bracket is a sharpening term, which can be used to prevent the diffusion of the surfactant on both side of the interface \cite{Jain2024JCP}.  

To obtain the equilibrium solution of the surfactant concentration equation, we consider the one-dimensional problem to be in a steady state with $\mathbf{u}=\mathbf{0}$. In this case, Eq. (\ref{eq-Jain0}) can be simplified by
\begin{equation}
	\frac{\mathrm{d}^2\psi}{\mathrm{d}x^2}-\frac{\mathrm{d}}{\mathrm{d}x}\left(\frac{\psi}{\delta}\frac{\mathrm{d}\delta}{\mathrm{d}x}\right)=0.
\end{equation}
By integrating the above equation and using the boundary condition $\mathrm{d}\psi/\mathrm{d}x\rightarrow0$ when $x\rightarrow\pm\infty$, we have
\begin{equation}
	\frac{1}{\psi}\frac{\mathrm{d}\psi}{\mathrm{d}x}=\frac{1}{\delta}\frac{\mathrm{d}\delta}{\mathrm{d}x}.
\end{equation}
From above equation, one can easily obtain $\psi=c_1\delta$ with $c_1$ being a constant, similar result is also reported in Ref. \cite{Yamashita2024JCP}. Based on the equilibrium profile of order parameter [see Eq. (\ref{eq-phieq})], the steady solution of surfactant concentration can be given by 
\begin{equation}
	\psi_{eq}=c_1\delta|_{\phi=\phi_{eq}}=c_1\left|\nabla\phi_{eq}\right|=c_1\frac{\phi_A-\phi_B}{W}\left[1-\tanh^2\left(\frac{2x}{W}\right)\right].
\end{equation}
To determine the constant $c_1$ in above expression, the condition $\psi=\psi_0$ at $x=0$ is considered, and then one can obtain the following steady distribution of the surfactant concentration,
\begin{equation}\label{eq-psieq}
	\psi_{eq}=\psi_0\left[1-\tanh^2\left(\frac{2x}{W}\right)\right].
\end{equation}
From Eq. (\ref{eq-psieq}) one can see that, as mentioned in Ref. \cite{Jain2024JCP}, the diffuse-interface model (\ref{eq-Jain}) is consistent with the phase-field model (\ref{eq-ACE}), which results in the transport of surfactant concentration along the interface without any unphysical numerical leakage into either of the phases on both sides of the interface. It is worth noting that the idea of preventing leakage of the surfactant concentration comes from the previous work on the transport of scalars in the bulk of one of two phases \cite{Jain2023JCP}. 
\subsection{Navier-Stokes equations for incompressible fluid flows}
To describe the fluid flow, the following consistent and conservative Navier-Stokes equations are applied \cite{Mirjalili2021JCP,Zhan2022PRE},
\begin{subequations}
    \begin{equation}
        \nabla\cdot\mathbf{u}=0, 
    \end{equation}
    \begin{equation}
        \frac{\partial\left(\rho\mathbf{u}\right)}{\partial t}+\nabla\cdot\left[\left(\rho\mathbf{u}-\mathbf{S}\right)\mathbf{u}\right]=-\nabla P+\nabla\cdot\mu\left[\nabla\mathbf{u}+\left(\nabla\mathbf{u}\right)^\top\right]+\mathbf{F}_s+\mathbf{F}_b,
    \end{equation}
\end{subequations}
where $\rho$ is the density, $P$ is the pressure, $\mu$ is the dynamic viscosity, $\mathbf{F}_s$ represents the surface tension force, and $\mathbf{F}_b$ is the body force. 
In phase-field method, the density and the viscosity of the fluid are usually assumed to be the linear functions of the order parameter,
\begin{equation}
    \rho=\frac{\phi-\phi_B}{\phi_A-\phi_B}\left(\rho_A-\rho_B\right)+\rho_B,\quad\mu=\frac{\phi-\phi_B}{\phi_A-\phi_B}\left(\mu_A-\mu_B\right)+\mu_B,
\end{equation}
where $\rho_k$ and $\mu_k$ are the density and viscosity of the pure fluid $k$ ($k=$A, B). $\mathbf{S}$ is the mass flux between different phases, and for the Allen-Cahn equation (\ref{eq-ACE}) adopted in this work, it can be given by \cite{Mirjalili2021JCP,Zhan2022PRE}
\begin{equation}
	\mathbf{S}=M_{\phi}\frac{\rho_A-\rho_B}{\phi_A-\phi_B}\left[\nabla\phi-\frac{4\left(\phi_A-\phi\right)\left(\phi-\phi_B\right)}{W\left(\phi_A-\phi_B\right)}\frac{\nabla\phi}{\left|\nabla\phi\right|}\right].
\end{equation}

When the surfactant is considered, the surface tension coefficient $\sigma$ is not a constant, and the following general continuous form of the surface tension force should be adopted \cite{Kim2005JCP},
\begin{equation}
    \mathbf{F}_s=\delta\left[-\sigma\left(\nabla\cdot\mathbf{n}\right)\mathbf{n}+\nabla_s\sigma\right].
\end{equation}
In phase-field method, the surface tension force can be further written as \cite{Liu2013PRE}
\begin{equation}
    \begin{aligned}
        \mathbf{F}_s=&\frac{3W}{2\left(\phi_A-\phi_B\right)^2}\nabla\cdot\sigma\left(\left|\nabla\phi\right|^2\mathbf{I}-\nabla\phi\nabla\phi\right)\\
        =&\frac{3\sigma}{2\left(\phi_A-\phi_B\right)^2}\left[\frac{16}{W}\left(\phi_A-\phi\right)\left(\phi-\phi_B\right)\left(\phi_A+\phi_B-2\phi\right)-W\nabla^2\phi\right]\nabla\phi\\
        &+\frac{3W}{2\left(\phi_A-\phi_B\right)^2}\left[\left|\nabla\phi\right|^2\nabla\sigma-\left(\nabla\sigma\cdot\nabla\phi\right)\nabla\phi\right],
    \end{aligned}
\end{equation}
where the first term is the capillary force, and the second term is the well-known Marangoni force. 

In the presence of the surfactant, an equation of state is required to build the relationship between the surface tension coefficient and the concentration of surfactant. Here the following Langmuir equation of state is considered,
\begin{equation}\label{eq-EOS}
	\sigma\left(\psi\right)=\sigma_0\left[1+E_0\ln\left(1-\psi\right)\right],
\end{equation}
where $\sigma_0$ is the surface tension coefficient of a clean interface where $\psi=0$, $E_0$ is the elasticity number which measures the sensitivity of $\sigma$ to the variation of $\psi$. 
When the surfactant concentration $\psi$ is small, the following linear approximation can be derived from Eq. (\ref{eq-EOS}),
\begin{equation}\label{eq-EOSlinear}
	\sigma\left(\psi\right)=\sigma_0\left(1-E_0\psi\right).
\end{equation} 

\section{Lattice Boltzmann method}\label{sec-LBM}
In this section, we will develop the phase-field-based LB method for the transport of insoluble surfactant in two-phase flows. Under the general multiple-relaxation-time LB framework \cite{Chai2020PRE, Chai2023PRE}, the specific LB models for different fields are presented as follows.
\subsection{Lattice Boltzmann model for phase field}
For the convection-diffusion type equation with an extra flux term, the evolution equation of LB model can be written as
\begin{equation}\label{eq-LBEphi}
    %f_i\left(\mathbf{x}+\mathbf{c}_i\Delta t,t+\Delta t\right)=f_i\left(\mathbf{x},t\right)-\Lambda_{ij}^f\left[f_j\left(\mathbf{x},t\right)-f_j^{eq}\left(\mathbf{x},t\right)\right]+\Delta t\left(\delta_{ij}-\frac{\Lambda_{ij}^f}{2}\right)F_j\left(\mathbf{x},t\right),
    f_i\left(\mathbf{x}+\mathbf{c}_i\Delta t,t+\Delta t\right)=f_i\left(\mathbf{x},t\right)-\left(\mathbf{M}^{-1}\mathbf{S}^f\mathbf{M}\right)_{ij}\left[f_j\left(\mathbf{x},t\right)-f_j^{eq}\left(\mathbf{x},t\right)\right]+\Delta t\left[\mathbf{M}^{-1}\left(\mathbf{I}-\mathbf{S}^f/2\right)\mathbf{M}\right]_{ij}F_j\left(\mathbf{x},t\right),
\end{equation}
where $f_i\left(\mathbf{x},t\right)$ represents the distribution function of phase field at position $\mathbf{x}$ and time $t$ along the $i$-th direction of discrete velocity $\mathbf{c}_i$ ($i=0,1,\cdots,q-1$ with $q$ being the number of discrete velocity directions).
%$\Delta t$ is the time step, $\left(\Lambda_{ij}^f\right)$ is the collision matrix.
$\Delta t$ is the time step, $\mathbf{M}$ is a $q\times q$ transformation matrix, and $\mathbf{S}^f$ is the relaxation matrix.
$f_i^{eq}\left(\mathbf{x},t\right)$ and $F_i\left(\mathbf{x},t\right)$ are the equilibrium and source distribution functions, which are given by \cite{Liang2018PRE,Wang2019Capillarity,Zhan2022PRE}
\begin{equation}
    f_i^{eq}=\omega_i\phi\left(1+\frac{\mathbf{c}_i\cdot\mathbf{u}}{c_s^2}\right),\quad F_i=\omega_i\frac{\mathbf{c}_i\cdot\partial_t\left(\phi\mathbf{u}\right)}{c_s^2}+\omega_i\mathbf{c}_i\cdot\frac{4\left(\phi_A-\phi\right)\left(\phi-\phi_B\right)}{W\left(\phi_A-\phi_B\right)}\frac{\nabla\phi}{\left|\nabla\phi\right|}.
\end{equation}
Here $\omega_i$ is the weight coefficient, and $c_s$ represents the lattice sound speed.

To correctly recover the phase-field equation (\ref{eq-ACE}) from the evolution equation (\ref{eq-LBEphi}), the relaxation parameter $s_1^f$ in the relaxation matrix, corresponding to the first-order moment of distribution function, is determined by
\begin{equation}
    \frac{1}{s_1^f}=\frac{M_{\phi}}{c_s^2\Delta t}+\frac{1}{2},
\end{equation}
and the order parameter is computed by
\begin{equation}
	\phi=\sum_if_i.
\end{equation}
\subsection{Lattice Boltzmann model for surfactant concentration field}
In this part, to solve the surfactant concentration equation (\ref{eq-Jain}), the LB model for surfactant concentration is designed as
\begin{equation}
%    g_i\left(\mathbf{x}+\mathbf{c}_i\Delta t,t+\Delta t\right)=g_i\left(\mathbf{x},t\right)-\Lambda_{ij}^g\left[g_j\left(\mathbf{x},t\right)-g_j^{eq}\left(\mathbf{x},t\right)\right]+\Delta t\left(\delta_{ij}-\frac{\Lambda_{ij}^g}{2}\right)G_j\left(\mathbf{x},t\right),
    g_i\left(\mathbf{x}+\mathbf{c}_i\Delta t,t+\Delta t\right)=g_i\left(\mathbf{x},t\right)-\left(\mathbf{M}^{-1}\mathbf{S}^g\mathbf{M}\right)_{ij}\left[g_j\left(\mathbf{x},t\right)-g_j^{eq}\left(\mathbf{x},t\right)\right]+\Delta t\left[\mathbf{M}^{-1}\left(\mathbf{I}-\mathbf{S}^g/2\right)\mathbf{M}\right]_{ij}G_j\left(\mathbf{x},t\right),
\end{equation}
where $\mathbf{S}^g$ is the relaxation matrix, $g_i\left(\mathbf{x},t\right)$ denotes the distribution function of concentration field, and other distribution functions appeared in above equation can be expressed as
\begin{equation}\label{eq-LBEpsi}
    g_i^{eq}=\omega_i\psi\left(1+\frac{\mathbf{c}_i\cdot\mathbf{u}}{c_s^2}\right),\quad G_i=\omega_i\frac{\mathbf{c}_i\cdot\partial_t\left(\psi\mathbf{u}\right)}{c_s^2}+\omega_i\mathbf{c}_i\cdot\frac{4\psi\left(\phi_A+\phi_B-2\phi\right)}{W\left(\phi_A-\phi_B\right)}\frac{\nabla\phi}{\left|\nabla\phi\right|}.
\end{equation}
The macroscopic concentration of surfactant is calculated by
\begin{equation}
    \psi=\sum_ig_i,
\end{equation}
and the relaxation parameter corresponding to the first-order moment of distribution function can be given by
\begin{equation}\label{eq-s1g}
    \frac{1}{s_1^g}=\frac{D}{c_s^2\Delta t}+\frac{1}{2}.
\end{equation}
It should be noted that above LB model can correctly recover the transport equation of insoluble surfactant (\ref{eq-Jain}) through the Chapman-Enskog or direct Taylor expansion, and some details are shown in \ref{sec-app}. 
\subsection{Lattice Boltzmann model for flow field}
Similar to the previous works \cite{Zhan2022PRE,Wang2019Capillarity}, the LB evolution equation for incompressible Navier-Stokes equations reads 
\begin{equation}
%    h_i\left(\mathbf{x}+\mathbf{c}_i\Delta t,t+\Delta t\right)=h_i\left(\mathbf{x},t\right)-\Lambda_{ij}^h\left[h_j\left(\mathbf{x},t\right)-h_j^{eq}\left(\mathbf{x},t\right)\right]+\Delta t\left(\delta_{ij}-\frac{\Lambda_{ij}^h}{2}\right)H_j\left(\mathbf{x},t\right),
    h_i\left(\mathbf{x}+\mathbf{c}_i\Delta t,t+\Delta t\right)=h_i\left(\mathbf{x},t\right)-\left(\mathbf{M}^{-1}\mathbf{S}^h\mathbf{M}\right)_{ij}\left[h_j\left(\mathbf{x},t\right)-h_j^{eq}\left(\mathbf{x},t\right)\right]+\Delta t\left[\mathbf{M}^{-1}\left(\mathbf{I}-\mathbf{S}^h/2\right)\mathbf{M}\right]_{ij}H_j\left(\mathbf{x},t\right),
\end{equation}
where $\mathbf{S}^h$ is also the relaxation matrix, the equilibrium and force distribution functions are designed as \cite{Zhan2022PRE}
\begin{equation}
    h_i^{eq}=\lambda_i+\omega_i\left[\frac{\mathbf{c}_i\cdot\rho\mathbf{u}}{c_s^2}+\frac{\left(\rho\mathbf{u}-\mathbf{S}\right)\mathbf{u}:\left(\mathbf{c}_i\mathbf{c}_i-c_s^2\mathbf{I}\right)}{2c_s^4}\right],\quad H_i=\omega_i\left[\mathbf{u}\cdot\nabla\rho+\frac{\mathbf{c}_i\cdot\mathbf{F}}{c_s^2}+\frac{\mathbf{m}_{2H}:\left(\mathbf{c}_i\mathbf{c}_i-c_s^2\mathbf{I}\right)}{2c_s^4}\right],
\end{equation}
with $\lambda_0=\rho_0+\left(\omega_0-1\right)P/c_s^2$, $\lambda_i=\omega_iP/c_s^2$ ($i\neq0$). In above equation, the moment $\mathbf{m}_{2H}$ is given by
\begin{equation}
    \mathbf{m}_{2H}=\partial_t\left(\rho\mathbf{uu}-\frac{\mathbf{Su}+\mathbf{uS}}{2}\right)+c_s^2\left(\mathbf{u}\nabla\rho+\nabla\rho\mathbf{u}\right).
\end{equation}

Finally, the macroscopic velocity and pressure are calculated by
\begin{subequations}
    \begin{equation}
        \mathbf{u}=\sum_i\mathbf{c}_ih_i+\frac{\Delta t}{2}\mathbf{F},
    \end{equation}
    \begin{equation}\label{eq-P}
        P=\frac{c_s^2}{1-\omega_0}\left[\sum_{i\neq0}h_i+\frac{\Delta t}{2}\mathbf{u}\cdot\nabla\rho-\omega_0\frac{\left(\rho\mathbf{u}-\mathbf{S}\right)\cdot\mathbf{u}}{2c_s^2}+K\frac{\Delta t}{2}\partial_t\left(\rho\mathbf{u}\cdot\mathbf{u}-\mathbf{S}\cdot\mathbf{u}\right)\right],
    \end{equation}
\end{subequations}
where $K$ is a parameter related to a specific discrete velocity structure, and the relaxation parameter corresponding to the second-order moment of distribution function can be determined by 
\begin{equation}
    \frac{1}{s_2^h}=\frac{\mu}{\rho c_s^2\Delta t}+\frac{1}{2}.
\end{equation}
\section{Numerical validations and discussion}\label{sec-simulations}
To validate the developed phase-field-based LB method for the transport of insoluble surfactant in two-phase flows, several two-dimensional benchmark tests are considered in this section. In the LB simulations, the D2Q9 lattice structure is applied, in this case, the parameter $K$ in Eq. (\ref{eq-P}) can be given by $K=\omega_0/\left(c^2-c_s^2\right)$, where $c_s^2=c^2/3$, $c=\Delta x/\Delta t$ with $\Delta x$ being the lattice spacing, and additionally, the half-way bounce-back scheme \cite{Ladd1994JFM} is applied to treat the no-flux and velocity boundary conditions. 
%Moreover, to improve the efficiency of the LB algorithm, the collision step is conducted in the moment space by a transformation between collision matrix $\bm{\Lambda}^j$ and the relaxation matrix $\mathbf{S}^j$ ($j=f,g,h$) through the relation $\bm{\Lambda}^j=\mathbf{M}^{-1}\mathbf{S}^j\mathbf{M}$. 
Moreover, to improve the efficiency of the LB algorithm, the collision step is conducted in the moment space by a transformation between moments $\mathbf{m}^k$ ($k=f,g,h$) and distribution functions through the relation $\mathbf{m}^k=\mathbf{M}\mathbf{f}^k$, where $\mathbf{f}^k$ represents a vector composed of the distribution functions. 
In this work, the following natural transformation matrix $\mathbf{M}$ and diagonal relaxation matrix $\mathbf{S}^k$ are used,
\begin{equation}
	\mathbf{M}=\begin{pmatrix}
		1 &  1 &  1 &  1 &  1 &  1 &  1 &  1 &  1\\
		0 &  1 &  0 & -1 &  0 &  1 & -1 & -1 &  1\\
		0 &  0 &  1 &  0 & -1 &  1 &  1 & -1 & -1\\
		0 &  1 &  0 &  1 &  0 &  1 &  1 &  1 &  1\\
		0 &  0 &  1 &  0 &  1 &  1 &  1 &  1 &  1\\		
		0 &  0 &  0 &  0 &  0 &  1 & -1 &  1 & -1\\	
		0 &  0 &  0 &  0 &  0 &  1 & -1 & -1 &  1\\	
		0 &  0 &  0 &  0 &  0 &  1 &  1 & -1 & -1\\
		0 &  0 &  0 &  0 &  0 &  1 &  1 &  1 &  1\\
	\end{pmatrix},\quad\mathbf{S}^k=\mathbf{diag}\left(s_0^k,s_1^k,s_1^k,s_2^k,s_2^k,s_2^k,s_3^k,s_3^k,s_4^k\right).
\end{equation}
\subsection{Surface diffusion of surfactant}\label{sec-diffusion}
We first consider the problem of the surface diffusion of surfactant to test the accuracy of the present LB model for surfactant concentration equation. The surfactant is non-uniformly distributed on the interface of a stationary droplet with radius $R$, and the concentration would diffuse along the interface until it reaches a steady state. Initially, the interfacial concentration can be set as $\hat{\psi}\left(\theta\right)=\left(1-\cos\theta\right)/2$ in polar coordinates, and the analytical solution of this problem can be expressed as \cite{Teigen2009CMS}
\begin{equation}
	\hat{\psi}\left(\theta,t\right)=\frac{1}{2}\left(1-e^{-\frac{D}{R^2}t}\cos\theta\right).
\end{equation}

\begin{figure}
	\centering
	\subfigure[]{
		\begin{minipage}{0.4\linewidth}
			\centering
			\includegraphics[width=2.0in]{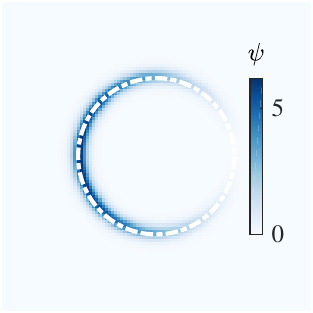}
	\end{minipage}}
	\subfigure[]{
		\begin{minipage}{0.4\linewidth}
			\centering
			\includegraphics[width=2.0in]{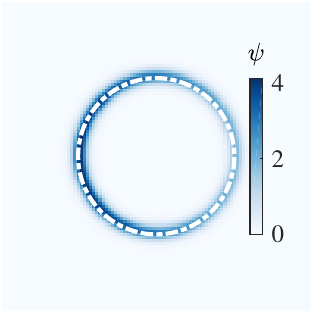}
	\end{minipage}}
	\caption{The surfactant diffusion on the surface of a stationary droplet [(a) $t=0$, (b) $t=1$, the white dashed line represents the fluid interface].}
	\label{fig-diffusionT}
\end{figure}

\begin{figure}
	\centering
	\includegraphics[width=3.5in]{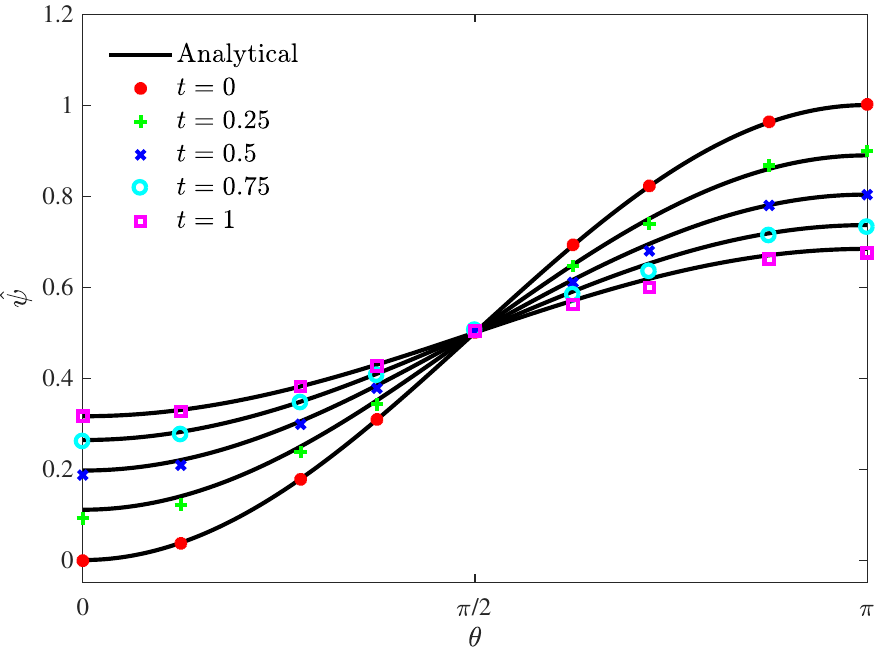}
	\caption{The distribution of surfactant concentration $\hat{\psi}$ along the interface at different times.}
	\label{fig-diffusionCom}
\end{figure}

In our simulations, the droplet with $R=1$ is placed in the center of the domain $\left[-2,2\right]\times\left[-2,2\right]$ with the periodic boundary condition imposed on all boundaries, and the physical parameters are set as $M_{\phi}=0.1$, $D=1$, $W=0.16$. We perform some simulations with $\Delta x=0.04$ and $\Delta t=0.005$, and present the distributions of the surfactant concentration at $t=0$ and $t=1$ in Fig. \ref{fig-diffusionT}. From this figure, one can observe that the surfactant diffuses along the interface without any artificial leakage into the bulk regions. To give a quantitative comparison, we plot the interfacial concentration versus the angle at different times in Fig. \ref{fig-diffusionCom}, and find that the present results are in a good agreement with the analytical solutions.

\subsection{Convection of a droplet}\label{sec-convetion}
\begin{figure}
	\centering
	\subfigure[$t=0$]{
		\begin{minipage}{0.4\linewidth}
			\centering
			\includegraphics[width=2.0in]{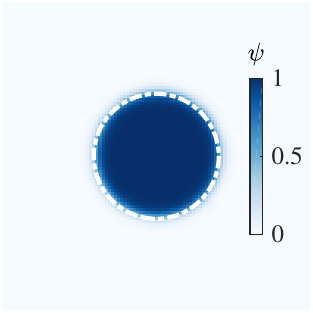}
	\end{minipage}}
	\subfigure[$t=2$]{
		\begin{minipage}{0.4\linewidth}
			\centering
			\includegraphics[width=2.0in]{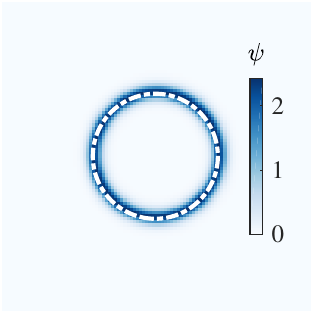}
	\end{minipage}}
	\caption{The convection of a droplet along with surfactant, the white dashed line denotes the fluid interface.}
	\label{fig-convectionT}
\end{figure}

\begin{figure}
	\centering
	\subfigure[$t=0$]{
		\begin{minipage}{0.48\linewidth}
			\centering
			\includegraphics[width=3.0in]{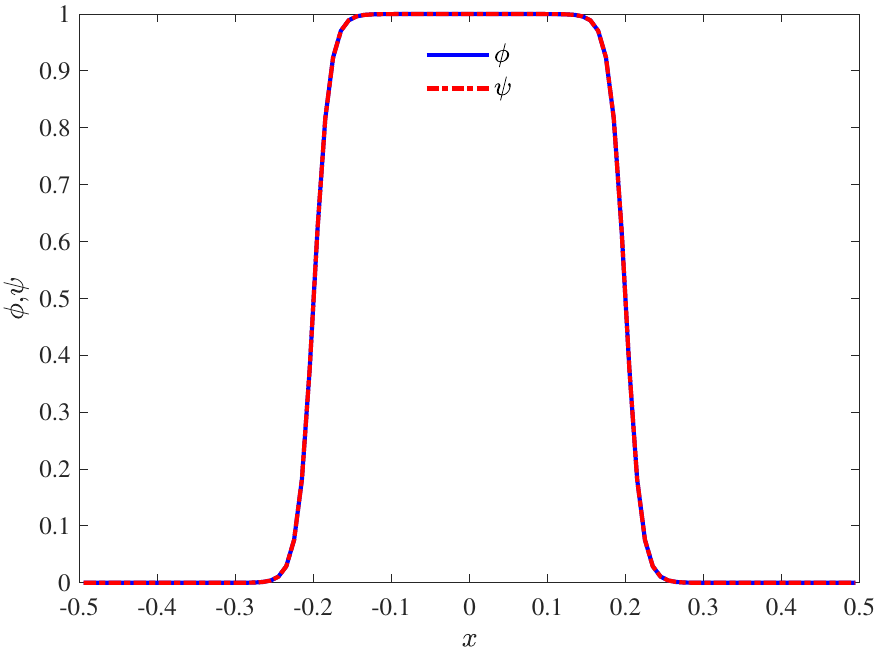}
	\end{minipage}}
	\subfigure[$t=2$]{
		\begin{minipage}{0.48\linewidth}
			\centering
			\includegraphics[width=3.0in]{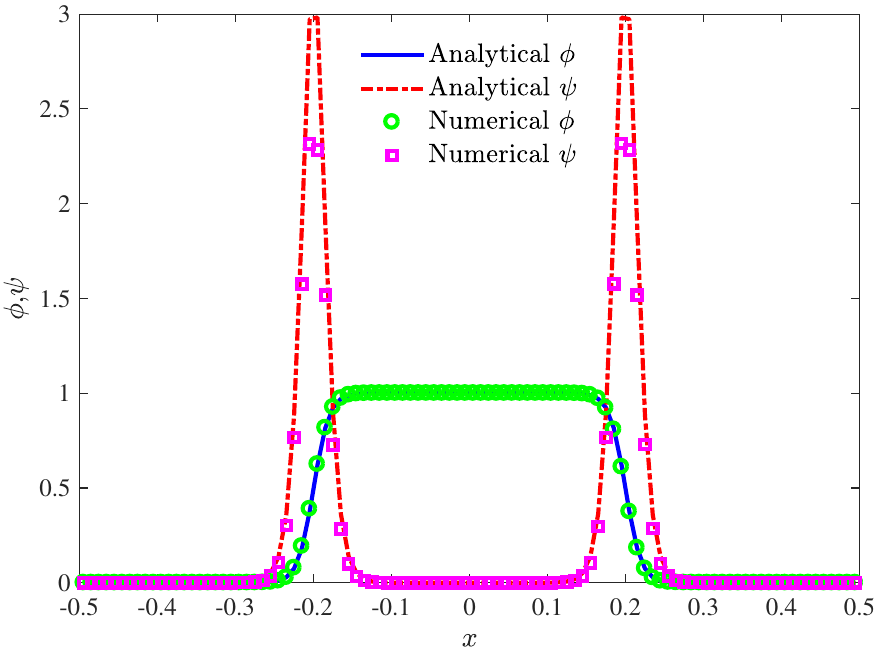}
	\end{minipage}}
	\caption{The profiles of fluid interface and surfactant concentration along $y=0$.}
	\label{fig-convectionP}
\end{figure}

In this part, the surfactant concentration field coupled with phase field is considered under the convection effect. A droplet with the radius $R=0.2$ is initially placed in the center of the computational domain $\left[-0.5,0.5\right]\times\left[-0.5,0.5\right]$ with the periodic boundary condition imposed on all boundaries, the concentration is initialized as $\psi=\phi$, and a uniform velocity $\mathbf{u}=\left(0.5,0\right)$ is imposed on phase and concentration fields. We conduct some simulations with $M_{\phi}=0.1$, $D=0.2$, $W=4\Delta x$, and plot the concentration field in Fig. \ref{fig-convectionT} where $\Delta x=0.01$ and $\Delta t=0.001$. Figure \ref{fig-convectionP} shows the profiles of the concentration and order parameter along $y=0.5$ at $t=0$ and $t=2$. From Figs. \ref{fig-convectionT} and \ref{fig-convectionP}, one can see that the surfactant reorganizes and moves to the interface region, which is consistent with the analytical solution. Here we also note that the maximum packing of concentration is related to the initial definition of surfactant concentration and the width of the fluid interface.

\subsection{A single rising bubble}\label{sec-bubble}
Now we focus on the problem of a single rising bubble, which can be considered as a fully coupled system. Initially, a circular bubble with radius $R$ is located at $\left(0,2R\right)$ in the rectangular domain $\left[-2R,2R\right]\times\left[0,8R\right]$. The no-slip boundary condition is imposed on top and bottom boundaries, and the periodic boundary condition is applied in $x$ direction. 

\begin{figure}
	\centering
	\subfigure[$t=0$]{
		\begin{minipage}{0.24\linewidth}
			\centering
			\includegraphics[width=1.5in]{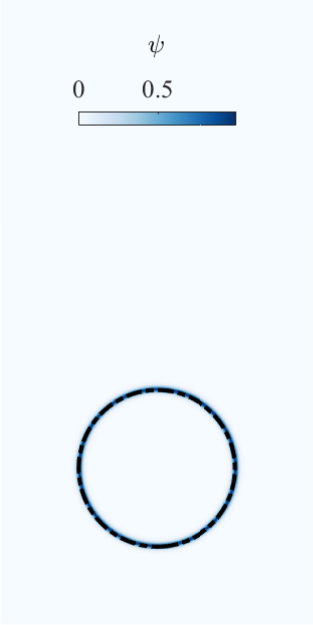}
	\end{minipage}}
	\subfigure[$t=1$]{
		\begin{minipage}{0.24\linewidth}
			\centering
			\includegraphics[width=1.5in]{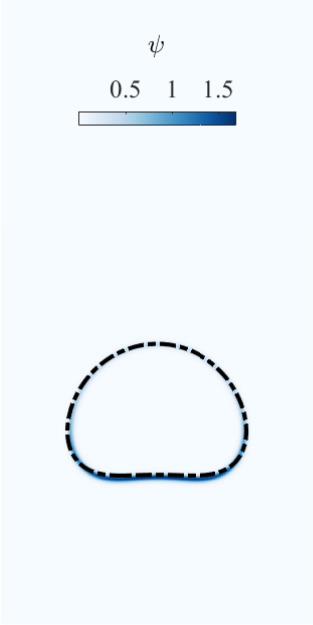}
	\end{minipage}}
	\subfigure[$t=2$]{
		\begin{minipage}{0.24\linewidth}
			\centering
			\includegraphics[width=1.5in]{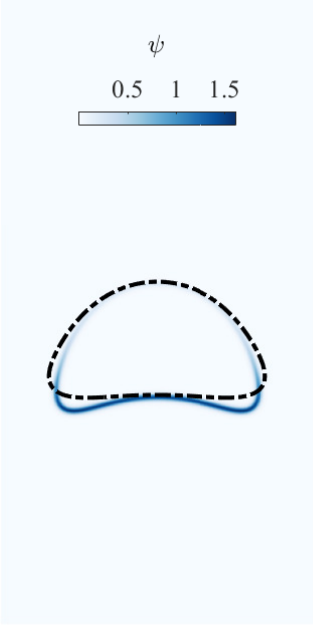}
	\end{minipage}}
	\subfigure[$t=3$]{
		\begin{minipage}{0.24\linewidth}
			\centering
			\includegraphics[width=1.5in]{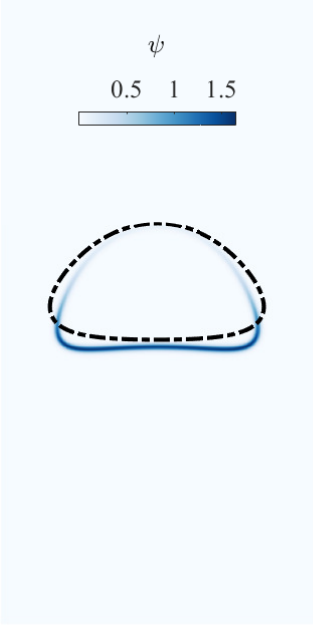}
	\end{minipage}}
	%\caption{The surfactant concentration at different times for $x_{in}=0.01$ with the black dashed line being the bubble shape in case of a clean interface ($x_{in}=0$).}
	\caption{The evolutions of bubble shape and surfactant concentration of two cases with $E_0=0$ and $E_0=0.5$ [the black dashed line represents the bubble shape of the case with $E_0=0$].}
	\label{fig-bubble}
\end{figure}

\begin{figure}
	\centering
%	\subfigure[Circularity]{
%		\begin{minipage}{0.48\linewidth}
%			\centering
%			\includegraphics[width=3.0in]{fig-bubble-Cp.pdf}
%	\end{minipage}}
%	\subfigure[Center of mass]{
%		\begin{minipage}{0.48\linewidth}
%			\centering
%			\includegraphics[width=3.0in]{fig-bubble-Yc.pdf}
%	\end{minipage}}
	\includegraphics[width=3.5in]{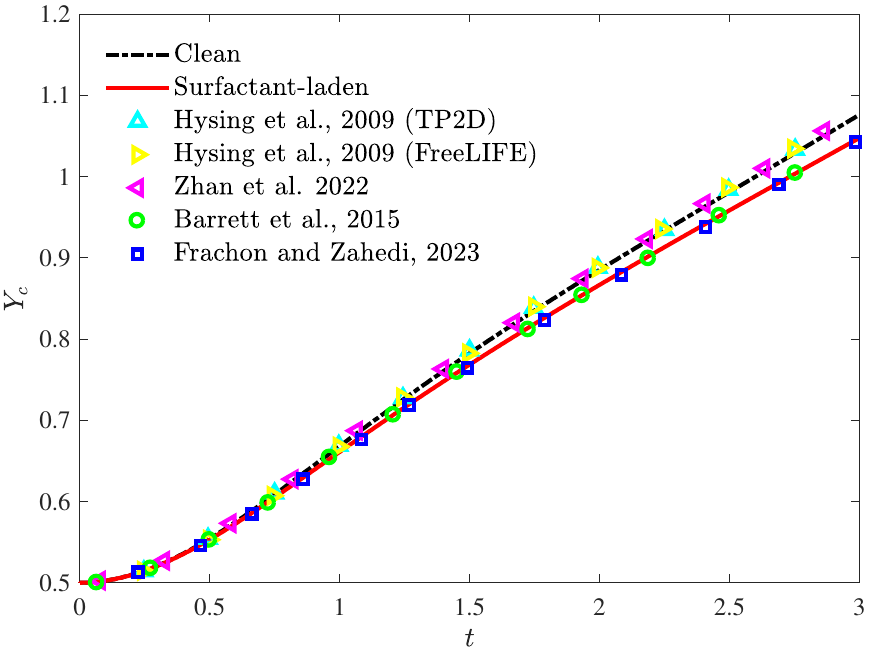}
	\caption{The center mass of single rising bubble.}
	\label{fig-bubbleCom}
\end{figure}

The dynamics of the single rising bubble can be characterized by the following Reynolds number, surface Péclet number and the Bond number,
\begin{equation}
	\mathrm{Re}=\frac{2\rho_AR\sqrt{2gR}}{\mu_A},\quad \mathrm{Pe}=\frac{2R\sqrt{2gR}}{D},\quad \mathrm{Bo}=\frac{4\rho_AgR^2}{\sigma_0},
\end{equation}   
where $g$ is the magnitude of the gravitational acceleration. 
%Here two values of $x_{in}$ are adopted, i.e., $x_{in}=0$ corresponding to the clean bubble and $x_{in}=0.01$ corresponding to the surfactant-laden bubble with the initial surfactant concentration $\psi=\left|\nabla\phi\right|$. 
Here the linear equation of state (\ref{eq-EOSlinear}) is applied and two values of $E_0$ are considered, i.e., $E_0=0$ and $E_0=0.5$ corresponding to the clean and surfactant-laden bubbles, respectively. For the latter case, the surfactant concentration is initialized by Eq. (\ref{eq-psieq}) with the interface concentration $\psi_0=1$. 
To compare with the results reported in some available works \cite{Hysing2009IJNMF,Zhan2022PRE}, some physical parameters are set as $\mathrm{Re}=35$, $\mathrm{Pe}=10$, $\mathrm{Bo}=10$, $\rho_A/\rho_B=10$, $\mu_A/\mu_B=10$ and $R=0.25$. In the following simulations, the other parameters are fixed as $W=4\Delta x$, $\Delta x=1/240$ and $\Delta t=\Delta x^2$. We plot the evolutions of the bubble and the surfactant concentration in Fig. \ref{fig-bubble}. As shown in this figure, the surfactant accumulates on the bottom of the bubble and increases the deformation of the interface, compared to the clean bubble. To give a quantitative comparison between the clean and surfactant-laden cases, the center mass ($Y_c$) of the bubble is measured and plotted in Fig. \ref{fig-bubbleCom}. From this figure, one can observe that compared to the clean case, the center mass of the surfactant-laden bubble decreases, which is also consistent with those in Refs. \cite{Barrett2015ESAIM,Frachon2023JCP}. 


\subsection{Deformation of a droplet in the shear flow}\label{sec-shear}
\begin{figure}
	\centering
	\includegraphics[width=3.0in]{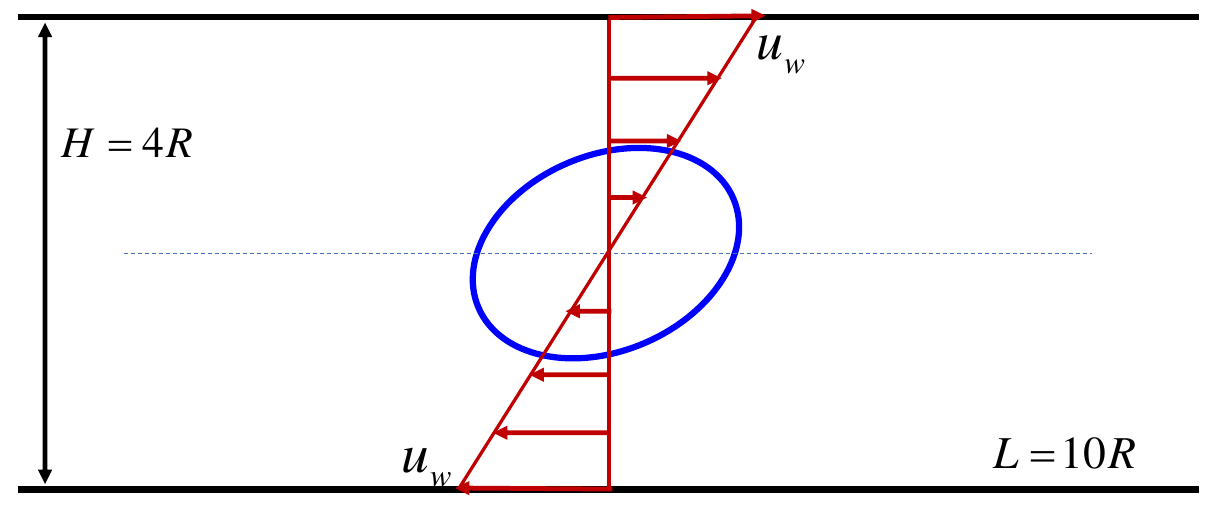}
	\caption{Schematic of a two-dimensional droplet in the shear flow.}
	\label{fig-shearInit}
\end{figure}

\begin{figure}
	\centering
	\subfigure[$E_0=0$]{
		\begin{minipage}{0.22\linewidth}
			\centering
			\includegraphics[width=1.4in]{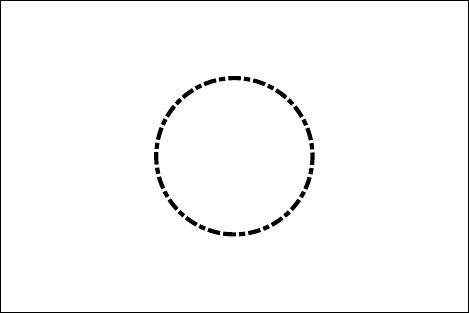}
		\end{minipage}
		\begin{minipage}{0.22\linewidth}
			\centering
			\includegraphics[width=1.4in]{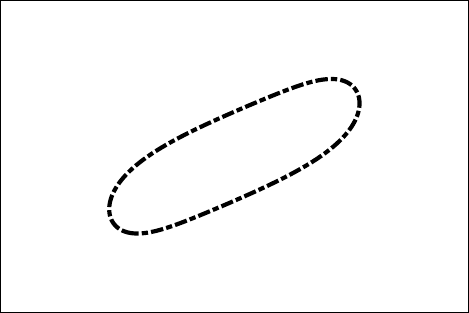}
		\end{minipage}
		\begin{minipage}{0.22\linewidth}
			\centering
			\includegraphics[width=1.4in]{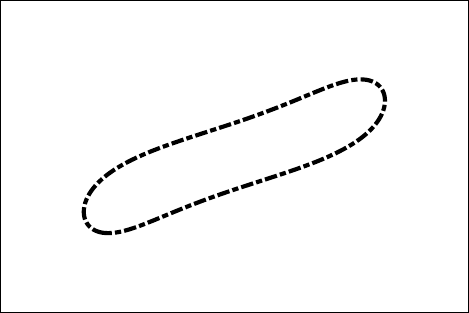}
		\end{minipage}
		\begin{minipage}{0.22\linewidth}
			\centering
			\includegraphics[width=1.4in]{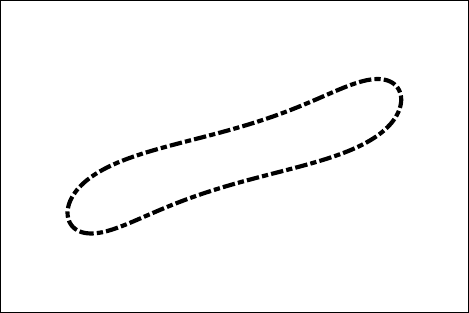}
	\end{minipage}}
	\subfigure[$E_0=0.25$]{
		\begin{minipage}{0.22\linewidth}
			\centering
			\includegraphics[width=1.4in]{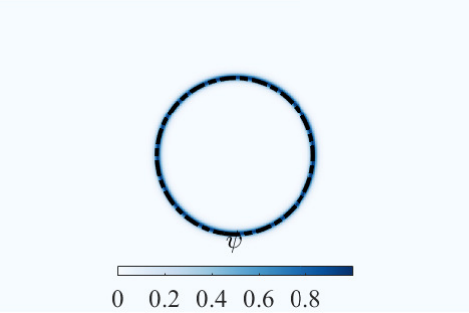}
		\end{minipage}
		\begin{minipage}{0.22\linewidth}
			\centering
			\includegraphics[width=1.4in]{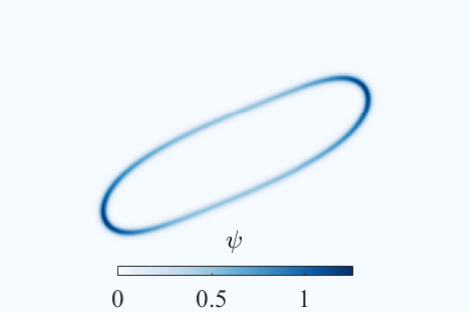}
		\end{minipage}
		\begin{minipage}{0.22\linewidth}
			\centering
			\includegraphics[width=1.4in]{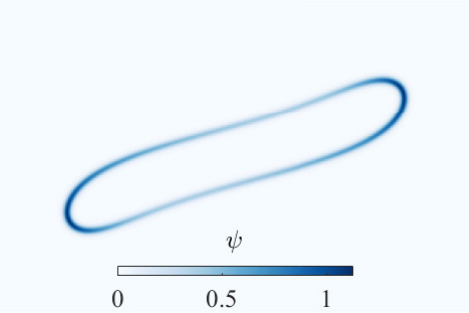}
		\end{minipage}
		\begin{minipage}{0.22\linewidth}
			\centering
			\includegraphics[width=1.4in]{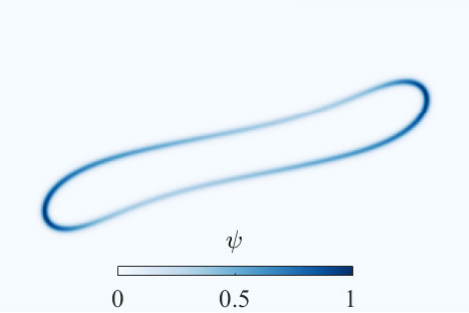}
	\end{minipage}}
	\subfigure[$E_0=0.5$]{
		\begin{minipage}{0.22\linewidth}
			\centering
			\includegraphics[width=1.4in]{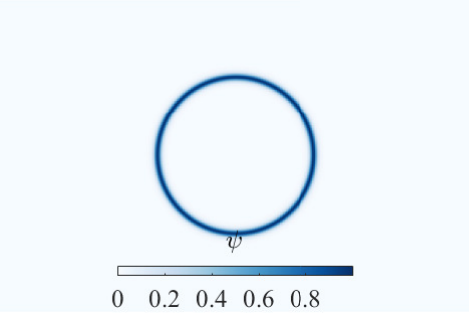}
		\end{minipage}
		\begin{minipage}{0.22\linewidth}
			\centering
			\includegraphics[width=1.4in]{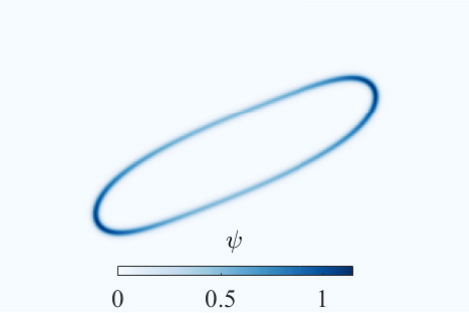}
		\end{minipage}
		\begin{minipage}{0.22\linewidth}
			\centering
			\includegraphics[width=1.4in]{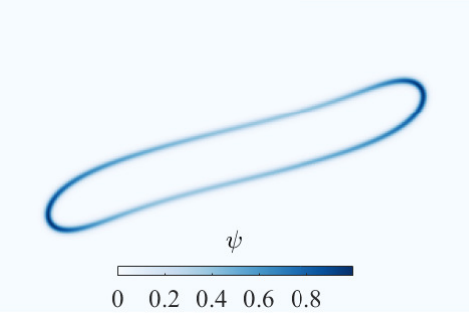}
		\end{minipage}
		\begin{minipage}{0.22\linewidth}
			\centering
			\includegraphics[width=1.4in]{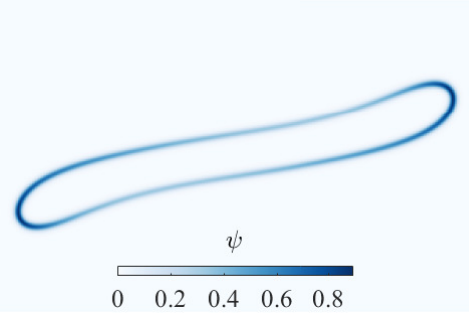}
	\end{minipage}}
	\caption{The droplet shape and surfactant concentration on the surface of a droplet in the shear flow [$t=0,4,8,12$ from left to right columns].}
	\label{fig-shear}
\end{figure}

\begin{figure}
	\centering
	\includegraphics[width=3.5in]{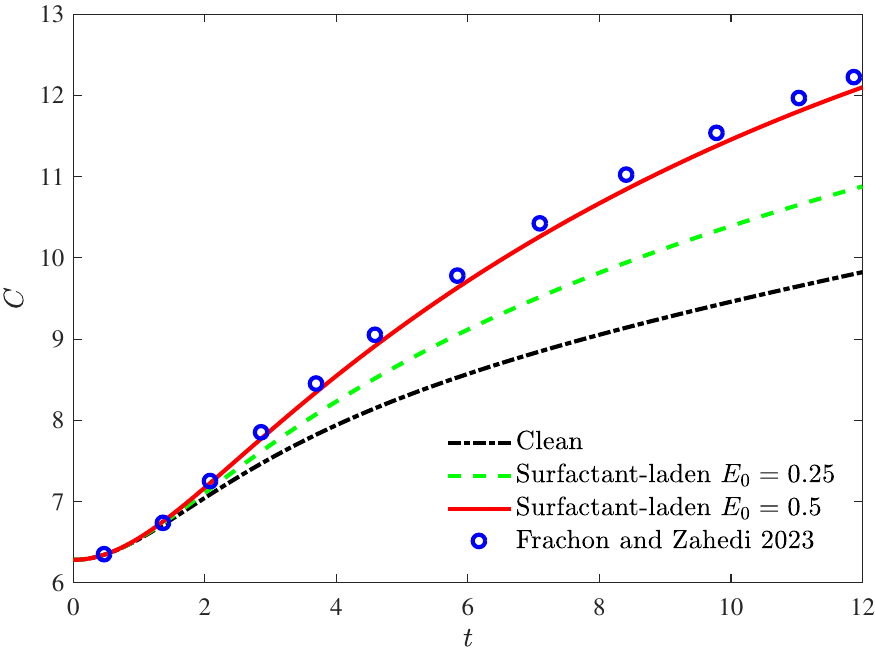}
	\caption{The total lengths of the droplet interface at different cases.}
	\label{fig-shearCom}
\end{figure}

The last problem we considered is the deformation of a droplet in the shear flow, which is also a benchmark problem in the study of the droplet dynamics in microfluidic applications. The schematic of this problem is shown in Fig. \ref{fig-shearInit} where a circular droplet with radius $R$ is initially placed in the center of the channel with the length $L=10R$ and width $H=4R$. The shear flow is induced by initial velocity $\mathbf{u}=\left(y/2,0\right)$ and the equal but opposite horizontal velocity velocities (the magnitude of velocity $u_w=1$) imposed on top and bottom walls. In our simulations, we set $\rho_A=\rho_B=1$, $\mu_A=\mu_B=0.1$, $\sigma_0=0.2$, and the equation of state and the initial concentration $\psi_0$ are the same as those in Sec. \ref{sec-bubble}. We conduct some simulations with three values of elasticity number, i.e., $E_0=0$, 0.25 and 0.5, and plot the shapes of droplet in Fig. \ref{fig-shear}, where $t=0,4,8,12$ from left to right columns. From this figure, one can find that with the increase of $E_0$, the droplet becomes more elongated and narrower. Additionally, a quantitative comparison of the lengths of the droplet interface ($C$) among different cases is also shown in Fig. \ref{fig-shearCom}, in which the length of the droplet interface is larger with a larger $E_0$, and it is also found that the present results also agree well with those in some previous works \cite{Frachon2023JCP,Barrett2015ESAIM}. These results clearly demonstrate the capacity of the present LB method for the interface dynamics of two-phase flows with insoluble surfactant.  


\section{Conclusions}\label{sec-conclusion}
In this paper, the second-order Allen-Cahn and diffuse-interface equations are considered to describe the transport of fluid interface and insoluble surfactant in two-phase flows, and compared to the classic Cahn-Hilliard phase-field models and surfactant transport models in the sharp representation \cite{Teigen2011JCP}, they are more consistent, much simpler and easier to be solved. 
The second-order model for the transport of insoluble surfactant is further extended to describe the insoluble surfactant two-phase flows with the Marangoni effect. Then the LB method is developed for the coupled system of phase, concentration, and flow fields, and compared to some hybrid numerical methods where the finite-difference schemes are adopted for the surfactant transport equation \cite{Liu2018JFM}, it is more efficient. 
%In this paper, a phase-field-based LB method is developed for the two-phase flows with insoluble surfactant. Different to the previous models based on Cahn-Hilliard framework, the second-order governing equations considered in this work are more simple and easy to be solved. In addition, the model proposed in Ref. \cite{Jain2024JCP} is extended to model the insoluble surfactant two-phase flows with Marangoni effect. 
To test the present LB method, the surface diffusion of surfactant on the surface of a droplet is first used to validate the LB model for surfactant concentration field, and then the system of coupled phase and surfactant concentration fields is further adopted. Finally, the fully coupled problems, a single rising bubble and the deformation of a droplet in the shear flow, are considered, and the results illustrate that the present LB method is accurate and efficient for the transport of the insoluble surfactant in two-phase flows.     

\section*{Acknowledgments}
One of the authors (Zhenhua Chai) thanks Porf. Yang Hu from Beijing Jiaotong University for useful discussion. This research was supported by the National Natural Science Foundation of China (Grants No. 12072127 and No. 123B2018), the Interdisciplinary Research Program of HUST (2024JCYJ001 and 2023JCYJ002), and the Fundamental Research Funds for the Central Universities, HUST (No. YCJJ20241101 and No. 2024JYCXJJ016). The computation was completed on the HPC Platform of Huazhong University of Science and Technology.

\appendix
\section{The direct Taylor expansion of present LB model for the transport equation of insoluble surfactant}\label{sec-app}
In this appendix, the direct Taylor expansion is adopted to recover the transport equation of insoluble surfactant from present LB model. Applying the Taylor expansion to Eq. (\ref{eq-LBEpsi}), we have
\begin{equation}
	\sum_{l=1}^N\frac{\Delta t^l}{l!}D_ig_i+O\left(\Delta t^{N+1}\right)=-\Lambda_{ij}^gg_j^{ne}+\Delta t\left(\delta_{ij}-\frac{\Lambda_{ij}^g}{2}\right)G_j,
\end{equation}
where $D_i=\partial_t+\mathbf{c}_i\cdot\nabla$, $\Lambda_{ij}^g=\left(\mathbf{M}^{-1}\mathbf{S}^g\mathbf{M}\right)_{ij}$, and $g_i^{ne}=g_i-g_i^{eq}$ is the non-equilibrium part of the distribution function $g_i$. From the above equation, one can obtain
\begin{subequations}
	\begin{equation}
		g_i^{ne}=O\left(\Delta t\right),
	\end{equation}
	\begin{equation}
		\sum_{l=1}^{N-1}\frac{\Delta t^l}{l!}D_i\left(g_i^{eq}+g_i^{ne}\right)+\frac{\Delta t^N}{N!}D_i^Ng_i^{eq}=-\Lambda_{ij}^gg_j^{ne}+\Delta t\left(\delta_{ij}-\frac{\Lambda_{ij}^g}{2}\right)G_j+O\left(\Delta t^{N+1}\right).
	\end{equation}
\end{subequations}
Then, we can derive equations at different orders of $\Delta t$,
\begin{subequations}
	\begin{equation}\label{eq-LB1order}
		D_ig_i^{eq}=-\frac{\Lambda_{ij}^g}{\Delta t}g_j^{ne}+\left(\delta_{ij}-\frac{\Lambda_{ij}^g}{2}\right)G_j+O\left(\Delta t\right),
	\end{equation}
	\begin{equation}\label{eq-LB2order0}
		D_i\left(g_i^{eq}+g_i^{ne}\right)+\frac{\Delta t}{2}D_i^2g_i^{eq}=-\frac{\Lambda_{ij}^g}{\Delta t}g_j^{ne}+\left(\delta_{ij}-\frac{\Lambda_{ij}^g}{2}\right)G_j+O\left(\Delta t^2\right).
	\end{equation}
\end{subequations}
From Eq. (\ref{eq-LB1order}), we get
\begin{equation}\label{eq-LB1orderD}
	\frac{\Delta t}{2}D_i^2g_i^{eq}=-\frac{1}{2}D_i\Lambda_{ij}^gg_j^{ne}+\frac{\Delta t}{2}D_i\left(\delta_{ij}-\frac{\Lambda_{ij}^g}{2}\right)G_j+O\left(\Delta t^2\right).
\end{equation}
Substituting Eq. (\ref{eq-LB1orderD}) into Eq. (\ref{eq-LB2order0}) yields
\begin{equation}\label{eq-LB2order}
	D_ig_i^{eq}+D_i\left(\delta_{ij}-\frac{\Lambda_{ij}^g}{2}\right)g_j^{ne}+\frac{\Delta t}{2}D_i\left(\delta_{ij}-\frac{\Lambda_{ij}^g}{2}\right)G_j=-\frac{\Lambda_{ij}^g}{\Delta t}g_j^{ne}+\left(\delta_{ij}-\frac{\Lambda_{ij}^g}{2}\right)G_j+O\left(\Delta t^2\right).
\end{equation}
To recover the correct governing equation (\ref{eq-Jain}), the collision matrix and distribution
functions should satisfy the following conditions,
\begin{subequations}
	\begin{equation}
		\sum_ie_i\Lambda_{ij}^g=s_0^gg_j,\quad\sum_i\mathbf{c}_i\Lambda_{ij}^g=s_1^g\mathbf{c}_j,
	\end{equation}
	\begin{equation}
		\sum_ig_i^{eq}=\psi,\quad\sum_I\mathbf{c}_ig_i^{eq}=\psi\mathbf{u},\quad\sum_i\mathbf{c}_i\mathbf{c}_ig_i^{eq}=\psi c_s^2\mathbf{I},
	\end{equation}
	\begin{equation}
		\sum_iG_i=0,\quad\sum_i\mathbf{c}_iG_i=\partial_t\left(\psi\mathbf{u}\right)+\frac{c_s^2}{D}\frac{4\psi\left(\phi_A+\phi_B-2\phi\right)}{W\left(\phi_A-\phi_B\right)}\frac{\nabla\phi}{\left|\nabla\phi\right|},
	\end{equation}
\end{subequations}
where $e_i=1$ for all $i=0,1,\cdots,q-1$.

Summing Eq. (\ref{eq-LB1order}) and Eq. (\ref{eq-LB2order}) over $i$, we can obtain
\begin{subequations}
	\begin{equation}
		\partial_t\psi+\nabla\cdot\left(\psi\mathbf{u}\right)=O\left(\Delta t\right),
	\end{equation}
	\begin{equation}\label{eq-sum2}
		\partial_t\psi+\nabla\cdot\left(\psi\mathbf{u}\right)+\nabla\cdot\left(1-\frac{s_1^g}{2}\right)\left[\sum_i\mathbf{c}_ig_i^{ne}+\frac{\Delta t}{2}\sum_i\mathbf{c}_iG_i\right]=O\left(\Delta t^2\right),
	\end{equation}
\end{subequations}
where $\sum_i\mathbf{c}_ig_i^{ne}$ can be obtained from Eq. (\ref{eq-LB1order}),
\begin{equation}\label{eq-cigi1}
	\sum_i\mathbf{c}_ig_i^{ne}=-\frac{\Delta t}{s_1^g}\left[\partial_t\left(\psi\mathbf{u}\right)+c_s^2\nabla\psi-\left(1-\frac{s_1^g}{2}\right)\sum_i\mathbf{c}_iG_i\right]+O\left(\Delta t^2\right).
\end{equation}
Substituting Eq. (\ref{eq-cigi1}) into Eq. (\ref{eq-sum2}), we can derive the transport equation of the insoluble surfactant (\ref{eq-Jain}) at the order of $O\left(\Delta t^2\right)$,
\begin{equation}
	\partial_t\psi+\nabla\cdot\left(\psi\mathbf{u}\right)=\nabla\cdot D\left[\nabla\psi-\frac{4\psi\left(\phi_A+\phi_B-2\phi\right)}{W\left(\phi_A-\phi_B\right)}\frac{\nabla\phi}{\left|\nabla\phi\right|}\right]+O\left(\Delta t^2\right),
\end{equation}
where the diffusivity $D$ satisfies the relation (\ref{eq-s1g}).

\bibliographystyle{elsarticle-num} 
\bibliography{references}

\begin{thebibliography}{10}
\expandafter\ifx\csname url\endcsname\relax
  \def\url#1{\texttt{#1}}\fi
\expandafter\ifx\csname urlprefix\endcsname\relax\def\urlprefix{URL }\fi
\expandafter\ifx\csname href\endcsname\relax
  \def\href#1#2{#2} \def\path#1{#1}\fi

\bibitem{Sun2014EF}
Q.~Sun, Z.~Li, S.~Li, L.~Jiang, J.~Wang, P.~Wang, Utilization of
  surfactant-stabilized foam for enhanced oil recovery by adding nanoparticles,
  Energy \& Fuels 28~(4) (2014) 2384--2394.

\bibitem{Eggleton2001PRL}
C.~D. Eggleton, T.-M. Tsai, K.~J. Stebe, Tip streaming from a drop in the
  presence of surfactants, Physical Review Letters 87 (2001) 048302.

\bibitem{Booty2005JFM}
M.~R. Booty, M.~Siegel, Steady deformation and tip-streaming of a slender
  bubble with surfactant in an extensional flow, Journal of Fluid Mechanics 544
  (2005) 243–275.

\bibitem{Baret2012LC}
J.-C. Baret, Surfactants in droplet-based microfluidics, Lab Chip 12 (2012)
  422--433.

\bibitem{Li2012WST}
Y.~Li, T.~Zhu, Y.~Liu, Y.~Tian, H.~Wang, {Effects of surfactant on bubble
  hydrodynamic behavior under flotation-related conditions in wastewater},
  Water Science and Technology 65~(6) (2012) 1060--1066.

\bibitem{Teigen2009CMS}
K.~E. Teigen, X.~Li, J.~Lowengrub, F.~Wang, A.~Voigt, A diffuse-interface
  approach for modelling transport, diffusion and adsorption/desorption of
  material quantities on a deformable interface, Communications in Mathematical
  Sciences 7~(4) (2009) 1009--1037.

\bibitem{Teigen2011JCP}
K.~{Erik Teigen}, P.~Song, J.~Lowengrub, A.~Voigt, A diffuse-interface method
  for two-phase flows with soluble surfactants, Journal of Computational
  Physics 230~(2) (2011) 375--393.

\bibitem{Stone1990PF}
H.~A. Stone, A simple derivation of the time‐dependent convective‐diffusion
  equation for surfactant transport along a deforming interface, Physics of
  Fluids A 2~(1) (1990) 111--112.

\bibitem{Xu2003JSC}
J.-J. Xu, H.-K. Zhao, An {Eulerian} formulation for solving partial
  differential equations along a moving interface, Journal of Scientific
  Computing 19 (2003) 573--594.

\bibitem{Liu2018JFM}
H.~Liu, Y.~Ba, L.~Wu, Z.~Li, G.~Xi, Y.~Zhang, A hybrid lattice {Boltzmann} and
  finite difference method for droplet dynamics with insoluble surfactants,
  Journal of Fluid Mechanics 837 (2018) 381–412.

\bibitem{Hu2021AML}
Y.~Hu, A diffuse interface–lattice {Boltzmann} model for surfactant transport
  on an interface, Applied Mathematics Letters 111 (2021) 106614.

\bibitem{Jain2024JCP}
S.~S. Jain, A model for transport of interface-confined scalars and insoluble
  surfactants in two-phase flows, Journal of Computational Physics 515 (2024)
  113277.

\bibitem{Yamashita2024JCP}
S.~Yamashita, S.~Matsushita, T.~Suekane, Conservative transport model for
  surfactant on the interface based on the phase-field method, Journal of
  Computational Physics 516 (2024) 113292.

\bibitem{Chen1998ARFM}
S.~Chen, G.~D. Doolen, Lattice {Boltzmann} method for fluid flows, Annual
  Reviews of Fluid Mechanics 30 (1998) 329--364.

\bibitem{Aidun2010ARFM}
C.~K. Aidun, J.~R. Clausen, Lattice-{Boltzmann} method for complex flows,
  Annual Review of Fluid Mechanics 42~(1) (2010) 439--472.

\bibitem{Chiu2011JCP}
P.-H. Chiu, Y.-T. Lin, A conservative phase field method for solving
  incompressible two-phase flows, Journal of Computational Physics 230~(1)
  (2011) 185--204.

\bibitem{Wang2019Capillarity}
H.~Wang, X.~Yuan, H.~Liang, Z.~Chai, B.~Shi, A brief review of the
  phase-field-based lattice {Boltzmann} method for multiphase flows,
  Capillarity 2~(2) (2019) 33--52.

\bibitem{Jain2023JCP}
S.~S. Jain, A.~Mani, A computational model for transport of immiscible scalars
  in two-phase flows, Journal of Computational Physics 476 (2023) 111843.

\bibitem{Mirjalili2021JCP}
S.~Mirjalili, A.~Mani, Consistent, energy-conserving momentum transport for
  simulations of two-phase flows using the phase field equations, Journal of
  Computational Physics 426 (2021) 109918.

\bibitem{Zhan2022PRE}
C.~Zhan, Z.~Chai, B.~Shi, Consistent and conservative phase-field-based lattice
  {Boltzmann} method for incompressible two-phase flows, Physical Review E 106
  (2022) 025319.

\bibitem{Kim2005JCP}
J.~Kim, A continuous surface tension force formulation for diffuse-interface
  models, Journal of Computational Physics 204~(2) (2005) 784--804.

\bibitem{Liu2013PRE}
H.~Liu, A.~J. Valocchi, Y.~Zhang, Q.~Kang, Phase-field-based lattice
  {Boltzmann} finite-difference model for simulating thermocapillary flows,
  Physical Review E 87 (2013) 013010.

\bibitem{Chai2020PRE}
Z.~Chai, B.~Shi, Multiple-relaxation-time lattice {Boltzmann} method for the
  {Navier-Stokes} and nonlinear convection-diffusion equations: {Modeling},
  analysis, and elements, Physical Review E 102 (2020) 023306.

\bibitem{Chai2023PRE}
Z.~Chai, X.~Yuan, B.~Shi, Rectangular multiple-relaxation-time lattice
  {Boltzmann} method for the {Navier-Stokes} and nonlinear convection-diffusion
  equations: {General} equilibrium and some important issues, Physical Review E
  108 (2023) 015304.

\bibitem{Liang2018PRE}
H.~Liang, J.~Xu, J.~Chen, H.~Wang, Z.~Chai, B.~Shi, Phase-field-based lattice
  {Boltzmann} modeling of large-density-ratio two-phase flows, Physical Review
  E 97 (2018) 033309.

\bibitem{Ladd1994JFM}
A.~J.~C. Ladd, Numerical simulations of particulate suspensions via a
  discretized {Boltzmann} equation. {Part} 1. {Theoretical} foundation, Journal
  of Fluid Mechanics 271 (1994) 285--309.

\bibitem{Hysing2009IJNMF}
S.~Hysing, S.~Turek, D.~Kuzmin, N.~Parolini, E.~Burman, S.~Ganesan, L.~Tobiska,
  Quantitative benchmark computations of two-dimensional bubble dynamics,
  International Journal for Numerical Methods in Fluids 60~(11) (2009)
  1259--1288.

\bibitem{Barrett2015ESAIM}
J.~W. Barrett, H.~Garcke, R.~Nürnberg, On the stable numerical approximation
  of two-phase flow with insoluble surfactant, ESAIM: M2AN 49~(2) (2015)
  421--458.

\bibitem{Frachon2023JCP}
T.~Frachon, S.~Zahedi, A cut finite element method for two-phase flows with
  insoluble surfactants, Journal of Computational Physics 473 (2023) 111734.

\end{thebibliography}
\end{document}